\documentclass[12pt]{article}
\usepackage{epsf}
\usepackage{graphics}

\begin{document}
\title
{Another origin of cosmological redshifts}
\author{Michael A. Ivanov \\
Physics Dept.,\\
Belarus State University of Informatics and Radioelectronics, \\
Minsk, Belarus\\ Email: ivanovma@gw.bsuir.unibel.by.}
\date{May 3, 2004}
\maketitle

\begin{abstract}
If gravitons are super-strong interacting particles which fulfill
a flat non-expanding universe, we would have another possibility
to explain cosmological redshifts - in a frame of non-kinematic
model. It is shown by the author that in this case SNe 1a data may
be understood without any dark energy and dark matter. A value of
relaxation factor is found in this paper. In this approach, we
have Newton's law of gravity as a simplest consequence, and the
connection between Newton's and Hubble's constants. A value of the
latter may be theoretically predicted.
\end{abstract}

\section[1]{Introduction }
If one assumes that the graviton background exists with the
Planckian spectrum and an effective temperature $T$, which we will
consider in a flat space-time, an energy of any photon decreases
with a distance $r,$ so a redshift $z$ is equal to \cite{1}:
$z=\exp(ar)-1.$ Here $a=H/c,$ where the Hubble constant $H=
(1/2\pi) D \cdot \bar \epsilon \cdot (\sigma T^{4}),$  $\bar
\epsilon$ is an average graviton energy, $\sigma$ is the
Stephan-Boltzmann constant, $D$ is a new constant. It is necessary
to have the following value of this constant: $D \sim 10^{-27}
m^{2}/eV^{2},$ i.e. gravitons should be super-strong interacting
particles. In this approach, the Newton constant $G$ is connected
with $H,$ that gives the following value of it \cite{2}: $H= 3.026
\cdot 10^{-18}s^{-1}=94.576 \ km \cdot s^{-1} \cdot Mpc^{-1}$ by
$T=2.7 K.$
\par
Additional photon flux's average energy losses on a way $dr$ due
to rejection of a part of photons from a source-observer direction
should be proportional to $badr,$ where the relaxation factor $b$
is equal to: $b=3/2 + 2/\pi = 2,137,$ as it is shown in the next
section.

\section[2]{How to calculate the factor $b$}
It is shown here how to find the value of relaxation factor $b$
which was used in author's paper \cite{1}. Let us assume that by
non-forehead collisions of a graviton with a photon, the latter
leaves a photon flux detected by a remote observer (an assumption
of narrow beam of rays). So as both particles have velocities $c,$
a cross-section of  interaction, which is "visible" under an angle
$\theta$ (see Fig. 1), will be equal to $\sigma_{0} \vert \cos
\theta \vert$ if $\sigma_{0}$ is a cross-section by forehead
collisions. The function $\vert \cos \theta \vert$ allows to take
into account both front and back hemispheres for riding gravitons.
Additionally, a graviton flux, which falls on a picked out area
(cross-section), depends on the angle $\theta.$ We have for the
ratio of fluxes:
$$\Phi(\theta)/\Phi_{0}=S_{s}/\sigma_{0}, $$
where $\Phi(\theta)$ and $\Phi_{0}$ are the fluxes which fall on
$\sigma_{0}$ under the angle $\theta$ and normally, $S_{s}$ is a
square of side surface of a truncated cone with a base
$\sigma_{0}$ (see Fig. 1).
\begin{figure}[th]
\epsfxsize=12.98cm \centerline{\epsfbox{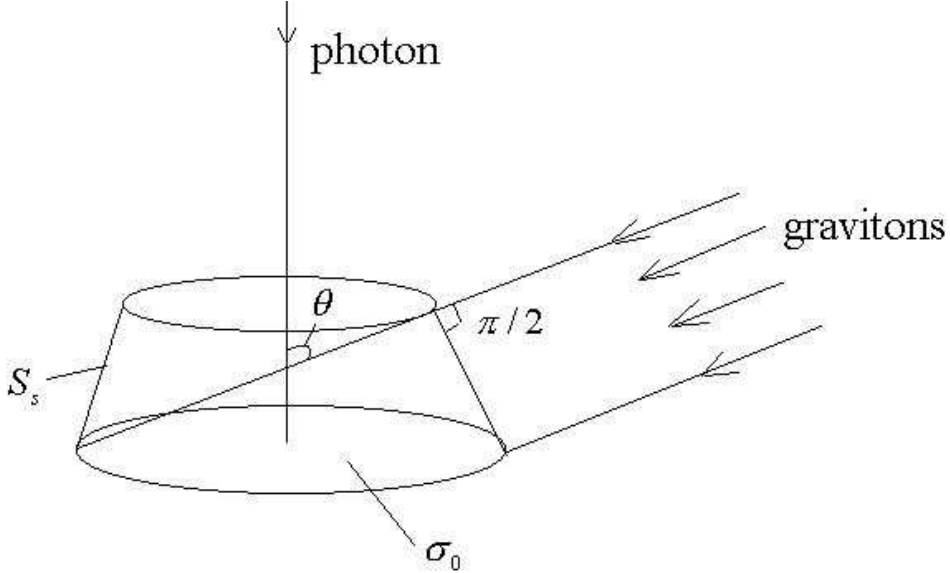}} \caption{By
non-forehead collisions of gravitons with a photon, it is
necessary to calculate a cone's side surface square, $S_{s}.$}
\end{figure}
Finally, we get for the factor $b:$
\begin{equation}
b=2 \int_{0}^{\pi/2}\cos{\theta}\cdot (S_{s}/\sigma_{0})\frac
{d\theta}{\pi/2}.
\end{equation}
By $0<\theta<\pi/4,$ a formed cone contains self-intersections,
and it is $S_{s}=2\sigma_{0} \cdot \cos{\theta}$. By
$\pi/4\leq\theta\leq\pi/2,$ we have $S_{s}=4\sigma_{0} \cdot
\sin^{2}{\theta}\cos{\theta}$.
\par
After computation of simple integrals, we get:
\begin{equation}
b=\frac {4}{\pi} (\int_{0}^{\pi/4}2\cos^{2}{\theta}d\theta +
\int_{\pi/4}^{\pi/2}\sin^{2}{2\theta}d\theta)= \frac {3}{2} +
\frac {2}{\pi} \simeq 2.137.
\end{equation}
\section[1]{Comparison with SNe 1a data}
This additional relaxation of any photonic flux due to
non-forehead collisions of gravitons with photons leads in this
model to the luminosity distance $D_{L}:$
\begin{equation}
D_{L}=a^{-1} \ln(1+z)\cdot (1+z)^{(1+b)/2} \equiv a^{-1}f_{1}(z).
\end{equation}
\par
\begin{figure}[th]
\epsfxsize=12.98cm \centerline{\epsfbox{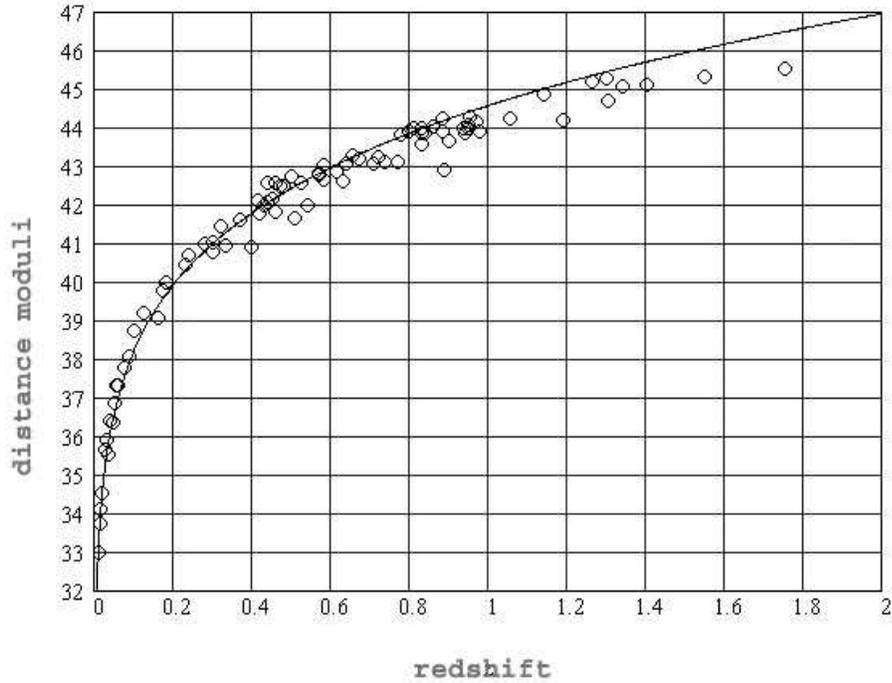}}
\caption{Comparison of the theoretical function $\mu_{0}(z)$
(solid line) with observations (points) from \cite{3} by Riess et
al.}
\end{figure}
The model may be compared with supernova data by Riess et al.
\cite{3} if one introduces distance moduli $\mu_{0} = 5 \log D_{L}
+ 25 = 5 \log f_{1} + c_{1}$, where $c_{1}$ is a constant (it is a
single free parameter to fit the data); $f_{1}$ is the luminosity
distance in units of $c/H$. In Figure 2, taken from my paper
\cite{5}, the function $\mu_{0}(z)$ is shown with $c_{1}=43$ to
fit observations for low redshifts; observational data are taken
from Table 5 of \cite{3}. The predictions fit observations very
well for roughly $z < 0.5$. Given this concordance between the
theory {\it without any kinematics} and observations for low
redshifts, we can think that any expansion of the universe {\it is
not necessary} for higher redshifts, too.
\par
There are discrepancies between predicted and observed values of
$\mu_{0}(z)$ for higher $z$. It would be explained in the model as
a result of deformation of SN spectra due to a discrete character
of photon energy losses. Today, a theory of this effect does not
exist, and its origin may be explained only qualitatively
\cite{4,5}.
\section[1]{Conclusion}
A main goal of any physical model is not to fit observational data
but to understand them to have a possibility to build an adequate
picture of the nature. There are a few facts - the Pioneer 10
anomaly \cite{6}, an existence of redshifts and SNe 1a specific
dimming, - which may be explained from one point of view in the
considered approach. This approach needs an existence of the
graviton background with very unexpected properties. Today, there
is not any model in physics of particles dealing with {\it an
external sea of particles} with a fixed spectrum. Our
understanding of cosmology, gravitational physics and physics of
particles would be essentially changed if this background really
exists.


\begin{thebibliography}{References  }
\bibitem{1}
M.A.Ivanov, General Relativity and Gravitation, {\bf 33}, 479
(2001); Erratum: {\bf 35}, 939 (2003); [astro-ph/0005084 v2].
\bibitem{2}
M.A.Ivanov. Screening the graviton background, graviton pairing,
and Newtonian gravity [gr-qc/0207006].
\bibitem{3}
A.G. Riess et al. Type Ia Supernova Discoveries at $z > 1$ From
the Hubble Space Telescope ..., [astro-ph/0402512] (to appear in
ApJ, 2004).
\bibitem{4}
Ivanov M.A. Contribution to QELS'95, May 21-26, 1995, Baltimore,
USA; Contribution to EQEC'96, Sept. 8-13, 1996, Hamburg, Germany.
\bibitem{5}
M.A.Ivanov. Another possible interpretation of SN 1a data -
without kinematics: Comments on the paper astro-ph/0402512 by A.
Riess et al. [astro-ph/0403130].
\bibitem{6}
J.D.Anderson et al. Phys. Rev. Lett., 1998, v.81, p. 2858; Phys.
Rev. {\bf D65} (2002) 082004. [gr-qc/0104064 v4]

\end{thebibliography}
\end{document}